\documentclass[11pt]{article}
\usepackage[margin=1in]{geometry}
\usepackage{amsmath,amssymb,bm}
\usepackage{graphicx}
\usepackage{booktabs}
\usepackage{caption}
\usepackage[backend=biber,style=phys,sorting=none]{biblatex}
\usepackage{hyperref}
\usepackage{xcolor}
\usepackage{microtype}
\usepackage{authblk}
\usepackage{setspace}
\usepackage[most]{tcolorbox}
\usepackage{enumitem}
\hypersetup{colorlinks=true,linkcolor=blue,citecolor=blue,urlcolor=blue}
\setlist[itemize]{leftmargin=1.4em,itemsep=0.15em,topsep=0.25em}

\definecolor{boxblue}{RGB}{242,247,252}
\definecolor{boxgreen}{RGB}{240,248,241}
\definecolor{boxred}{RGB}{252,243,243}

\title{Single Dirac Fermions on a Lattice:\\
A Unified View of Chiral and Parity Anomalies in Condensed Matter}
\author{Shun-Qing Shen}
\affil{Department of Physics, The University of Hong Kong}
\date{August 22, 2026}

\begin{document}
\maketitle

\begin{abstract}
Quantum anomalies are usually formulated in continuum field theory,
where regularization may violate a classical symmetry. In a crystal, the compact Brillouin zone and 
the high-energy bands make that ultraviolet completion physical. 
This Perspective develops a unified lattice view of chiral and parity 
anomalies through the problem of realizing a single massless Dirac fermion. 
The central motif is a finite 
low-energy region in which the relevant symmetry is restored, embedded in a high-energy sector that breaks the 
symmetry and completes the lattice theory. In $1+1$ and $3+1$ dimensions, 
the quantized chiral response is protected by local chiral symmetry near the Fermi surface. In $2+1$ dimensions, 
the half-integer Hall response belongs to a single massless Dirac band with mirror symmetry, 
not to a generic massive Dirac insulator that breaks the symmetry explicitly. 
In both cases the lattice perspective reveals how the low-energy symmetry emerges 
from the high-energy sector and protects the anomalous and quantized response. 
Recent experimental progress in quantum materials is discussed in light of this unified view.
\end{abstract}

\section{Introduction: from realizing anomalies to understanding them}
Noether's theorem provides the natural starting point for quantum anomalies \cite{Noether1918}. 
A continuous symmetry of a classical action $S$ implies a conservation law,
\begin{equation}
  \delta S=0 \quad \Longrightarrow \quad \partial_\mu j^\mu=0.
\end{equation}
For electrons, the most familiar example is the global $U(1)$ phase symmetry, which gives electric-charge conservation. 
The charge density and charge current form a four-vector $j^\mu=(\rho,\mathbf j)$ (in units with $c=1$) that satisfies the continuity equation 
$\partial_\mu j^\mu=0$. In a massless Dirac theory an additional chiral symmetry gives a conserved axial current 
at the classical level. Quantum anomalies expose a subtle limitation of this reasoning: 
after quantization, a regulator may be required that cannot preserve all classical symmetries simultaneously. 
The Adler--Bell--Jackiw (ABJ) anomaly is a canonical example 
\cite{Adler1969,BellJackiw1969,Fujikawa1979}.

A large literature now treats quantum materials as platforms in which field-theory anomalies can be realized and measured. 
Weyl and Dirac semimetals, topological insulators, Chern insulators, and related systems have been discussed 
in connection with chiral, parity, scale, and mixed anomalies 
\cite{NiemiSemenoff1983,Redlich1984,Redlich1984b,Armitage2018,Lv2021,Chernodub2022,Pettes2025}. 
This viewpoint has produced a productive dialogue between high-energy physics and condensed-matter physics. 
It has also generated a practical question that is more specific to solids: 
what does the regulator actually mean in a quantum material? 
How, then, can quantum anomalies be realized in a crystal, where the Brillouin zone is compact and the bands are periodic?

In a crystal, the ultraviolet completion is not an abstract prescription. The Brillouin zone is compact and periodic, 
the relevant lattice bands have finite bandwidth, and the high-energy states that complete a low-energy Dirac cone are part 
of the same band structure. This makes condensed matter more than a simulator of quantum field theory. 
It can serve as a microscope for the anomaly itself: the lattice shows where the regularization resides 
and how the low-energy symmetry is connected to the high-energy states.

\begin{figure}[t]
  \centering
  \includegraphics[angle=-90, width=0.95\linewidth]{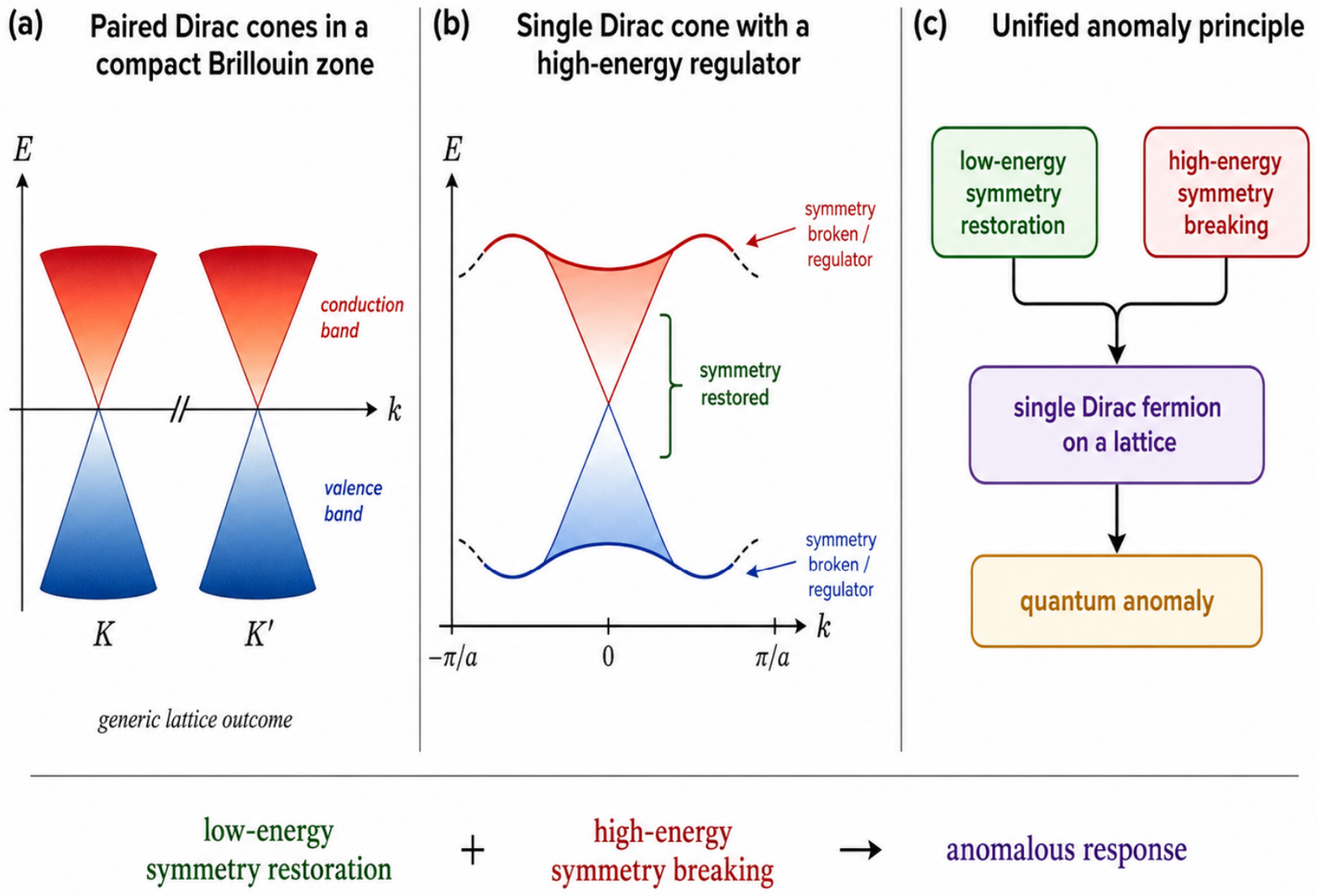}
  \caption{Single versus paired Dirac fermions in a compact Brillouin zone. Under the usual 
  assumptions behind fermion doubling, paired cones are the natural lattice outcome. 
  A single low-energy cone requires a high-energy completion that modifies the spectrum or the 
  protecting symmetry away from the Dirac region.}
  \label{fig:single-paired}
\end{figure}

The perspective developed here is therefore deliberately narrower than recent broad surveys of quantum anomalies 
in quantum materials \cite{Pettes2025}. We focus on a single organizing problem: 
how can a single Dirac fermion be consistently realized on a lattice? The answer reveals a common structure 
behind the chiral anomaly in odd spatial dimensions and the parity anomaly in even spatial dimensions. 
For the single-Dirac-fermion realizations considered here, the central principle is
\begin{equation}
\boxed{
\text{low-energy symmetry restoration}
+
\text{high-energy symmetry breaking}
\Longrightarrow
\text{anomalous response}.}
\label{eq:principle}
\end{equation}
The low-energy symmetry protects the anomalous and quantized response, whereas the high-energy symmetry-breaking sector 
supplies the physical regulator required by the compact Brillouin zone. The single-versus-paired lattice 
geometry is summarized schematically in Fig.~\ref{fig:single-paired}.

The discussion proceeds from this lattice question rather than from a catalogue of anomaly-related materials. 
We first connect a $1+1$-dimensional chiral anomaly to edge spectral flow and the quantum anomalous Hall effect, 
and then use the lowest Landau level to recover the $3+1$-dimensional ABJ response. 
The measured negative longitudinal magnetoresistivity in Weyl and Dirac semimetals 
is interpreted as a consistency check rather than a unique identifier of the anomaly.
We next revisit the parity anomaly in $2+1$ dimensions, emphasizing the
distinction between a massive Dirac fermion in a continuum regularization and
a single massless Dirac cone embedded in a lattice. This leads naturally to
the parity-anomalous semimetal and its characteristic combination of
half-quantized Hall response and finite longitudinal conductivity. Finally,
we discuss how these examples support a unified lattice perspective on
quantum anomalies and their realization in condensed matter systems.

\section{A single Dirac fermion in a compact Brillouin zone}

The Nielsen--Ninomiya theorem is the standard warning against treating an isolated continuum cone 
as a complete lattice theory \cite{NielsenNinomiya1981}. In one common formulation, 
locality, Hermiticity, translational invariance, and exact chiral symmetry 
over the full Brillouin zone cannot all be maintained while retaining a single unpaired massless fermion. 
The theorem is often introduced as an obstacle: it suggests that a single Dirac cone cannot be realized 
on a lattice without introducing additional degrees of freedom or breaking symmetries. 
Electrons on a graphene lattice provide a familiar example: 
the two inequivalent Dirac cones at the $K$ and $K'$ points are required by the lattice geometry.
For the present purpose it is more useful 
as a diagnostic: a single cone forces us to identify which assumption or condition 
is modified by the ultraviolet completion.

Wilson fermions provide the canonical lattice-gauge-theory answer \cite{Wilson1977,KarstenSmit1981}. 
A momentum-dependent term removes the unwanted low-energy doubler but breaks the ordinary continuum chiral or parity symmetry away from the selected node. Related lattice formulations show that this ultraviolet completion need not always take the form of an ordinary Wilson mass: the Ginsparg--Wilson construction retains a modified lattice chiral symmetry, while domain-wall fermions realize chiral modes as boundary zero modes \cite{GinspargWilson1982,Kaplan1992}. In lattice gauge theory these constructions are regulators whose continuum limit is the final target. In a crystal the interpretation is different. There is no instruction to throw away the high-energy sector: the states near the Brillouin-zone boundary are physical bands, and their symmetry character can enter measurable response functions.

\begin{figure}[t]
  \centering
  \includegraphics[angle=-90, width=0.94\linewidth]{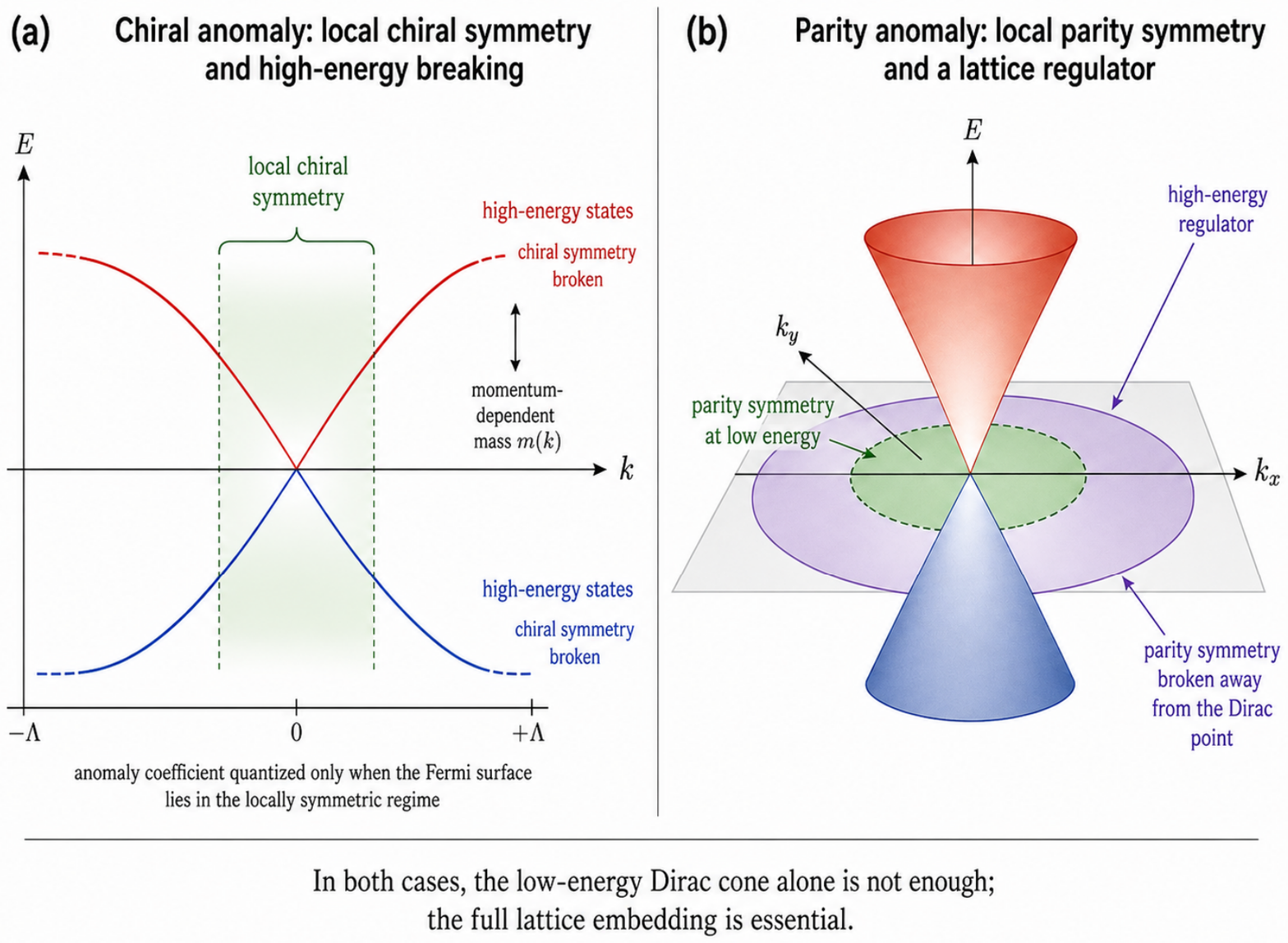}
  \caption{Local symmetry and high-energy regularization. (a) In the chiral problem, the anomaly coefficient is quantized while the Fermi surface remains inside a finite low-energy chiral-symmetric window; high-energy states break chiral symmetry and complete the lattice spectrum. (b) In the parity problem, the single massless Dirac cone is plotted with energy $E$ along the vertical axis and $(k_x,k_y)$ in the momentum plane. Parity is restored near the Dirac point and broken by the high-energy lattice regulator.}
  \label{fig:regulator}
\end{figure}

The quantum-anomalous semimetal program makes this distinction explicit \cite{FuNPJ2022,ZouPRB2022,ZouPRB2023,FuShen2025,ShenCoshare2024}. 
The important object is not only a Dirac point at $\mathbf k=0$. There can be a \emph{finite} momentum 
or energy window in which the low-energy states obey the relevant Dirac symmetry, 
while states outside that window provide the symmetry-breaking completion. 
The corresponding physical criterion is that the Fermi surface remain inside this locally 
symmetric region. This is stronger than the statement that the symmetry is recovered only 
in the mathematical limit $k\to0$. In odd spatial dimensions this logic is naturally associated with chiral symmetry, 
whereas in even spatial dimensions it can be associated with parity, 
which in two dimensions is an orientation-reversing mirror symmetry. Figure~\ref{fig:regulator} 
summarizes this finite low-energy symmetry window and its high-energy completion.

This finite symmetry window has a natural microscopic meaning for topological boundary states \cite{ZhangNatPhys2009,HasanKane2010,QiZhang2011}. 
A surface or edge Dirac state is sharply localized near the boundary of the system while its energy lies 
in the bulk band gap. 
At larger momenta it evolves continuously into bulk states and ceases to be an autonomous 
low-energy Dirac degree of freedom \cite{ShenSpringer2012}. The effective symmetry can therefore be exact or 
parametrically accurate throughout a finite range even though it is not a global symmetry 
of the complete band structure. In the chiral problem this construction was made explicit 
by deriving quasi-boundary Hamiltonians from higher-dimensional Chern insulators over the full Brillouin 
zone \cite{WangFuShen2025}. In the parity problem the same logic appears in semimagnetic topological 
insulators and in lattice models with a single Dirac cone \cite{ZouPRB2023,ShenPRB2026}.

There is an instructive contrast with the usual continuum language. 
In field theory the ultraviolet regulator is often introduced to define an otherwise divergent expression. 
Here the ultraviolet sector also tells us \emph{where} the emergent symmetry ceases to be valid. 
The anomaly is therefore not an isolated property of the linearized cone and not a purely ultraviolet artifact. 
It is an infrared--ultraviolet relation: the low-energy symmetry protects the quantized anomalous coefficient, 
while the high-energy completion makes the single-cone lattice theory possible.

\section{Chiral anomaly: spectral flow, Hall pumping, and the lattice regulator}

\subsection{$1+1$ dimensions and the boundary of a Chern insulator}

The $1+1$-dimensional problem gives the cleanest physical picture. 
Consider the boundary spectrum of a two-dimensional Chern insulator stripe. 
Two opposite edges host counterpropagating chiral modes.
Near the Fermi energy the edge mode is massless and chiral. At higher energy it cannot continue  
as an isolated one-dimensional linear branch; it bends and eventually merges into the two-dimensional bulk. 
The quasi-edge theory therefore contains a finite chiral-symmetric region connected 
to a high-energy chiral-symmetry-breaking sector \cite{WangFuShen2025}.

An electric field along the edge drives spectral flow. 
For a full massless Dirac theory the vector current remains conserved 
while the axial current acquires the familiar anomaly. 
In the lattice quasi-edge theory the same pumping can be written schematically as
\begin{equation}
  \partial_t\rho_5+\partial_x j_5
  = C_5(E_F)\frac{eE}{\pi\hbar},
  \label{eq:1d-anomaly}
\end{equation}
where the coefficient $C_5(E_F)$ depends on the chemical potential, or equivalently on the Fermi points $E_F$. 
It is quantized to be one while those Fermi points remain in the locally chiral-symmetric window 
and evolves once they enter the high-energy completion. 
In this case, the rates of change of the charge densities are
$d\rho_{R/L}/dt = \mp \frac{e^2}{h}E$,
which is equivalent to saying that the Hall current flows from one edge to the other. The total charge is conserved,
$d(\rho_{R}+\rho_{L})/dt =0$,
but the axial charge, or $\gamma^5$ charge, is not:
$d(\rho_{R}-\rho_{L})/dt =-2\frac{e^2}{h}E$.
This is a useful reversal of the usual intuition: the symmetry at the Fermi surface 
protects the quantized coefficient, but the spectral flow is completed by states far from the Fermi surface.

The same boundary picture also gives a direct way to understand the quantum anomalous Hall effect. 
A single chiral edge channel, considered by itself, has a boundary gauge anomaly: under an applied $E_x$ its charge appears not to be conserved within the one-dimensional theory. In the complete two-dimensional Chern insulator there is no violation of electric-charge conservation. The transverse bulk Hall current supplies exactly the missing flow at one edge and removes it at the opposite edge. In other words, the Hall current is the anomaly inflow required to make the anomalous boundary theory consistent.

For $C$ net chiral boundary channels this consistency condition gives the Hall conductance
$\sigma_{yx}=C\,\frac{e^2}{h}$.
This does not mean that the $1+1$-dimensional anomaly by itself creates a Chern insulator. 
The bulk Chern number establishes that the chiral edge modes exist; 
the anomaly-inflow argument explains why those modes cannot be autonomous 
and why their spectral flow is tied to the quantized transverse current. 
It is the same physics described from the viewpoints of TKNN topology, Laughlin pumping, 
bulk--edge correspondence, and boundary anomaly inflow \cite{TKNN1982,Laughlin1981,Hatsugai1993,CallanHarvey1985}.

\subsection{$3+1$ dimensions: many one-dimensional channels in the lowest Landau level}

The $3+1$-dimensional chiral anomaly can be understood as a direct extension
of the $1+1$-dimensional spectral-flow picture. A magnetic field $\mathbf B$ quantizes
the motion of charged particles normal to the field to form a series of Landau levels 
with the finite Landau degeneracy. 
At the lowest Landau level, an electric field parallel to $\mathbf B$ then drives spectral
flow along these effectively one-dimensional channels. The three-dimensional
anomaly can therefore be viewed as the same one-dimensional pumping process
repeated with the Landau degeneracy $n_L=\frac{eB}{h}$. Multiplying the pumping rate by the
number of flux quanta gives
\begin{equation}
  \partial_t\rho_5+\nabla\!\cdot\!\mathbf J_5
  = C_5\frac{e^2}{2\pi^2\hbar^2}\,
  \mathbf E\!\cdot\!\mathbf B .
  \label{eq:3d-anomaly}
\end{equation}
Again $C_5$ is a function of the chemical potential and Fermi surface topology. 
For an ideal continuum massless Dirac fermion, $C_5=1$.

\begin{figure}[t]
  \centering
  \includegraphics[angle=-90, width=0.98\linewidth]{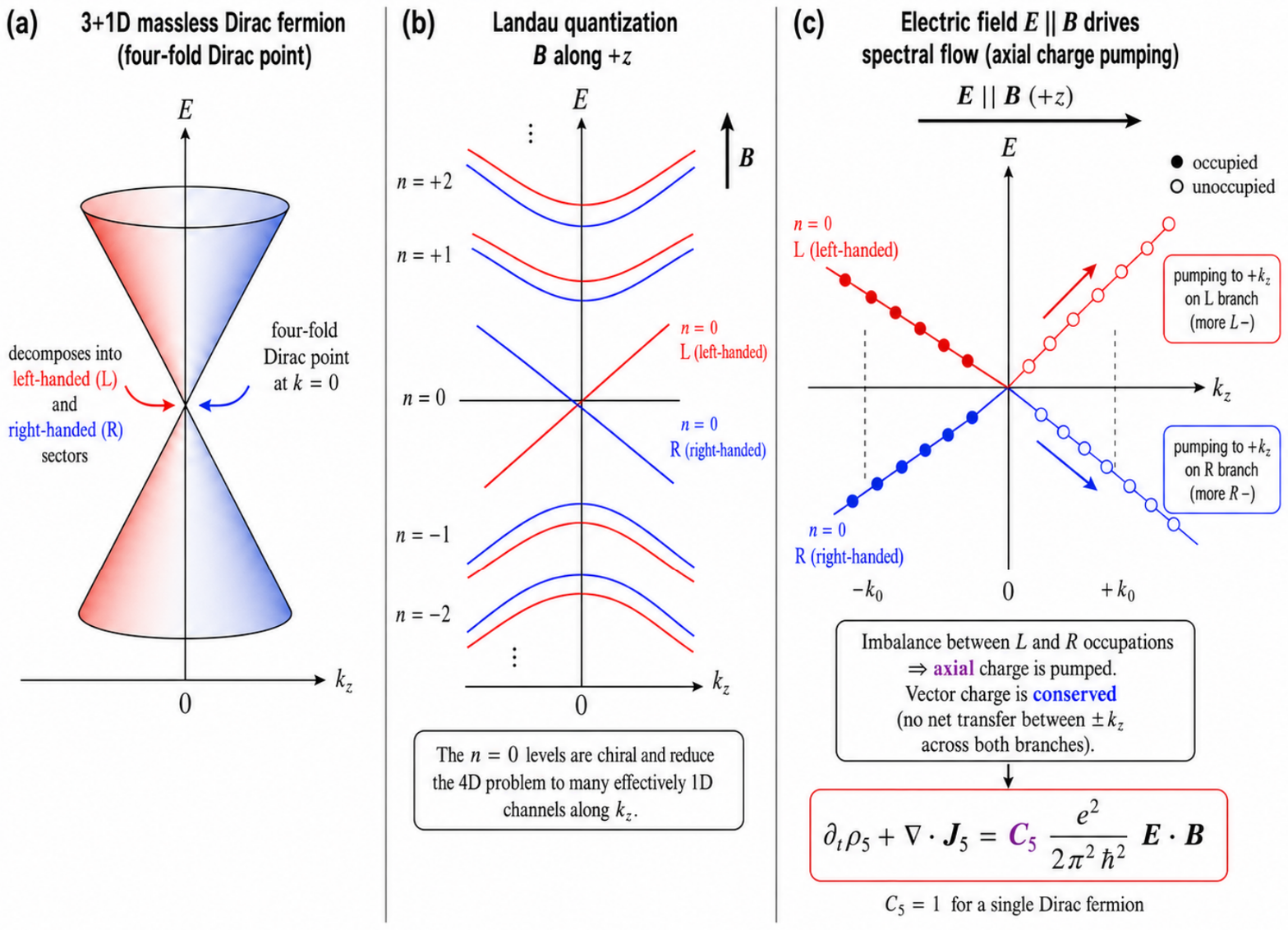}
  \caption{Landau-level picture of the $3+1$-dimensional chiral anomaly. A fourfold Dirac point at $\mathbf{k}=0$ decomposes into left- and right-handed sectors at the same momentum. For $B\parallel z$, the higher Landau levels ($|n|\ge 1$) remain dispersive in $k_z$, whereas the two $n=0$ branches are chiral. An electric field parallel to $\mathbf B$ drives spectral flow along these lowest-Landau-level channels and pumps axial charge while conserving total electric charge.}
  \label{fig:abj}
\end{figure}

On a lattice, however, $C_5=1$ is controlled by whether the physical Fermi
surface lies within the locally chiral-symmetric part of the spectrum
\cite{WangFuShen2025}. This can be made particularly transparent by embedding
the three-dimensional massless Dirac theory as the boundary theory of a
four-dimensional Chern insulator. In this construction, chiral symmetry is
preserved within a finite low-energy window inherited from the bulk gap,
whereas the high-energy states break chiral symmetry and provide the lattice
completion. If this symmetry-preserving window is sufficiently wide, a finite
Fermi surface can remain entirely inside it and the system exhibits the same
quantized pumping coefficient as a one-dimensional massless Dirac fermion.
Once the Fermi surface extends into the symmetry-breaking sector, $C_5$
deviates from unity. In the opposite limit, where the symmetry-preserving
window becomes very narrow, the construction approaches the conventional
Wilson-fermion limit: $C_5$ tends to unity only as the Fermi surface shrinks
toward the Dirac point. This distinction is largely invisible in the standard
continuum formulation, where the anomaly is usually discussed independently
of the finite carrier-density window that is unavoidable in a material.
The lowest-Landau-level reduction to many one-dimensional spectral-flow
channels is illustrated in Fig.~\ref{fig:abj}.

\subsection{Material realizations and the lattice viewpoint}
Dirac and Weyl semimetals provide the principal material platforms in which
the $3+1$-dimensional chiral anomaly has been explored \cite{NielsenNinomiya1983,SonSpivak2013,Armitage2018,OngLiang2021}. In Weyl semimetals
such as TaAs, parallel electric and magnetic fields produce spectral pumping
between Weyl nodes of opposite chirality, and negative longitudinal
magnetoresistance has been reported together with spectroscopic and
band-structure evidence for the Weyl states
\cite{ZhangNatComm2016}. In Dirac semimetals such as Cd$_3$As$_2$, a magnetic
field reorganizes the low-energy spectrum into Landau levels and produces
effectively chiral lowest-Landau-level channels. The observed negative
longitudinal magnetoresistance is particularly pronounced at low carrier
density, where the Fermi surface approaches the Dirac point
\cite{LiNatComm2016}. From the present lattice viewpoint, these material
realizations suggest a sharper criterion than the existence of a Dirac or
Weyl crossing alone: the anomalous response should be most robust when the
actual Fermi surface remains inside the finite low-energy region in which
chirality is well defined.

A finite Dirac mass changes the interpretation. Chirality is then explicitly
broken, while helicity can remain a useful band label in a magnetic field.
The resulting helical response crosses continuously to the chiral result at
high energy or in the massless limit, but is suppressed near the band edge
\cite{WangFuShen2021}. This distinction is important experimentally:
negative longitudinal magnetoresistivity in a gapped Dirac material need not
demonstrate the massless ABJ anomaly. More generally, 
longitudinal magnetotransport should be regarded as
consistency evidence rather than as a unique identifier of the chiral
anomaly. Current jetting, spatial inhomogeneity, weak-localization corrections,
mobility fluctuations, tensor mixing, and contact geometry can all produce or
enhance negative longitudinal magnetoresistance
\cite{ZhangNatComm2016,LiNatComm2016,OngLiang2021}.

For the present Perspective, the important point is methodological. A quantum
anomaly in a material should be identified through the combined evidence of
symmetry, band structure, and response, rather than from a single
magnetotransport curve. The central question is not merely whether a material
contains a Dirac or Weyl crossing, but whether its physical Fermi surface lies
inside the locally symmetric Dirac regime and whether the high-energy lattice
states provide the completion required for the anomalous low-energy theory.

\section{Parity anomaly: from the massive continuum cone to a massless lattice cone}

The parity anomaly is often presented through a massive $2+1$-dimensional Dirac fermion. 
That route correctly reveals the need for regularization, but it can obscure what is measurable 
in a lattice material. The first step is to state the symmetry precisely. 
In two spatial dimensions parity may be chosen as a mirror reflection, 
for example $(x,y)\mapsto(x,-y)$, not space reflection $(x,y)\mapsto(-x,-y)$. With an electric field $E_x$ along $x$, 
the perturbing potential is mirror even while the transverse current $J_y$ is mirror odd. 
An exactly mirror-symmetric system therefore cannot carry a Hall current. Unlike the continuous chiral symmetry relevant in $1+1$ and $3+1$ dimensions, parity in $2+1$ dimensions is a discrete symmetry.

A Dirac mass changes this immediately. For a two-component cone, the mass term is odd 
under the mirror and explicitly breaks parity. A nonzero Hall current then becomes symmetry allowed. 
This elementary point is useful because it separates \emph{explicit symmetry breaking} from \emph{anomaly}: 
the fact that a massive Dirac Hamiltonian permits Hall response is not, by itself, the parity anomaly.

\begin{tcolorbox}[colback=boxblue,colframe=blue!45!black,title={Mirror test for the Hall response},fonttitle=\bfseries]
For $\mathcal P_y:(x,y)\mapsto(x,-y)$ and $\mathbf E=E_x\hat x$, one has $E_x\to E_x$ but $J_y\to-J_y$. Exact mirror symmetry therefore requires $\sigma_{yx}=0$. A Dirac mass is mirror odd and makes $\sigma_{yx}\neq0$ symmetry allowed; it does not determine the value of $\sigma_{yx}$ and does not by itself establish an anomaly.
\end{tcolorbox}

The classic calculations of Niemi--Semenoff and Redlich remain essential \cite{NiemiSemenoff1983,Redlich1984,Redlich1984b}. 
Early condensed-matter formulations by Semenoff, Fradkin--Dagotto--Boyanovsky, and Haldane connected the same parity-odd Dirac physics to lattice systems \cite{Semenoff1984,FradkinDagottoBoyanovsky1986,Haldane1988}. A Wilson-fermion lattice regularization was analyzed explicitly by Coste and L\"uscher \cite{CosteLuscher1989}, and modern Hamiltonian and topological-insulator treatments further clarify the role of the regulator and the higher-dimensional bulk \cite{Lapa2019,MulliganBurnell2013}. 
These works show that a regularized continuum Dirac theory acquires a parity-odd Chern--Simons coefficient 
and expose the tension among gauge invariance, parity, and regularization. The condensed-matter question is narrower: 
can the resulting half-quantized coefficient be identified with the standalone dc Hall conductance 
of one massive two-dimensional lattice cone?

A finite Brillouin zone says no. For a massive continuum cone, the Berry-curvature integral is only 
exactly half quantized after taking an infinite momentum cutoff. 
A lattice cannot take that limit independently of its periodic band structure. 
Once the high-energy completion is included, a fully gapped translationally invariant 
two-dimensional band insulator obeys the TKNN integer quantization. 
A Wilson-type lattice term supplies the contribution missing from the isolated cone 
and produces either zero or an integer multiple of $e^2/h$ \cite{ShenPRB2026}. 
The massive cone and the massless cone therefore represent different physical limits, 
not two interchangeable descriptions of the same dc transport state.

Importantly, this conclusion does not invalidate the continuum effective action. 
It changes the condensed-matter interpretation. Half-cone contributions remain useful 
in low-energy descriptions of the Haldane model. 
What should be avoided is promoting such a contribution automatically to an independently 
measurable half-integer dc conductance of a fully gapped two-dimensional lattice subsystem, such as valley physics, magnetically gapped 
topological-insulator surfaces, axion electrodynamics, and layer-resolved response \cite{QiHughesZhang2008,EssinMooreVanderbilt2009}.

\begin{figure}[t]
  \centering
  \includegraphics[angle=-90, width=0.92\linewidth]{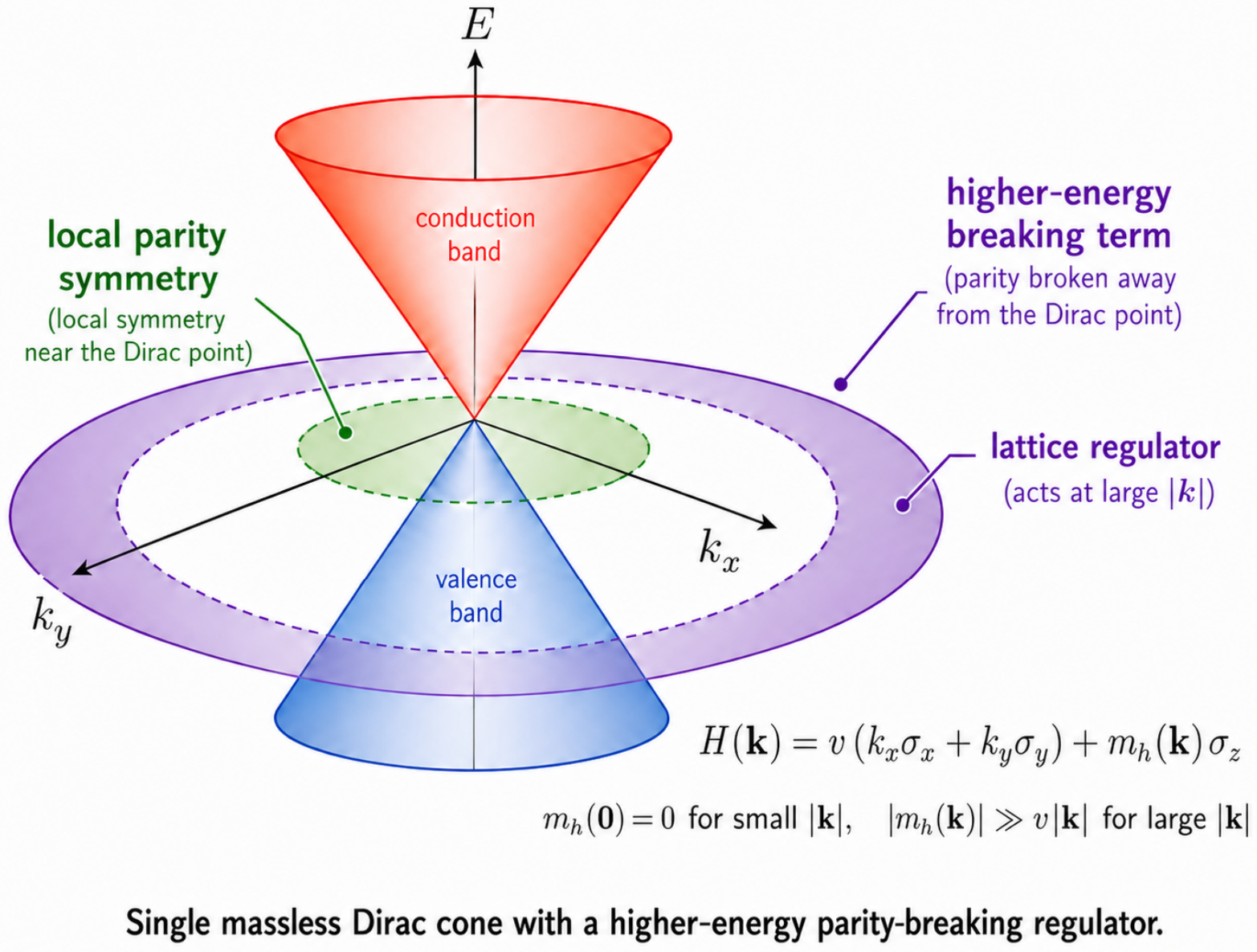}
  \caption{A single massless Dirac cone with a higher-energy parity-breaking regulator. Parity symmetry is restored near the Dirac point, whereas the regulator dominates at large momentum and completes the lattice theory.}
  \label{fig:massless-regulator}
\end{figure}

The correct lattice object for the half response is instead a single \emph{massless} cone 
whose parity symmetry is restored in a finite low-energy region 
and broken by the high-energy completion \cite{FuNPJ2022,ZouPRB2022,ZouPRB2023,FuShen2025,ShenPRB2026}. 
In idealized effective models the symmetry-breaking mass can be switched on only beyond a momentum scale $k_c$; 
in microscopic surface-state constructions the same crossover occurs as the Dirac surface state merges into bulk bands. 
Importantly, the quasiparticle spectrum can remain continuous and differentiable 
through the crossover even though the symmetry character changes. Figure~\ref{fig:massless-regulator} 
illustrates this separation between the low-energy massless cone and the high-energy parity-breaking regulator.


The characteristic transport signature is correspondingly
\begin{equation}
  \sigma_{yx}=\pm\frac{e^2}{2h},
  \qquad
  \sigma_{xx}>0 .
  \label{eq:pas-response}
\end{equation}
The sign of $\sigma_{yx}$ is determined by the part of the high energy band. The finite longitudinal conductivity is not a defect in the half-Hall state. It is part of its identity as a parity-anomalous semimetal (PAS). The same gaplessness also permits power-law rather than exponentially localized edge-current profiles \cite{ZouPRB2022}. This is the key conceptual shift: the parity anomaly in a lattice material belongs to an unpaired massless Dirac sector embedded in a parity-breaking ultraviolet completion.

\section{Half-quantized Hall effect: from semi-magnetic topological insulators to a field-tuned parity-anomalous semimetal}

The recent experimental development of magnetic topological insulators is
particularly instructive for the parity anomaly because it gradually turns
what began as a static materials realization into a controllable
single-Dirac-fermion problem. A three-dimensional topological-insulator thin
film naturally hosts two surface Dirac cones \cite{ZhangNatPhys2009,HasanKane2010,QiZhang2011}. In a symmetric film their
parity-odd responses compensate one another, whereas magnetic engineering
can make the two surfaces inequivalent and leave one low-energy Dirac sector
effectively unpaired. The central question is then not simply whether a
half-quantized Hall response can be observed, but which part of the complete
lattice system carries that response and how the single gapless cone is
completed at high energy.

The semi-magnetic topological insulator studied by Mogi \emph{et al.}
provided an important first realization of this situation
\cite{Mogi2022}. Only one surface was magnetically doped and gapped, while
the opposite surface remained nonmagnetic and gapless. A half-quantized Hall
response was observed without Landau-level quantization. The conventional
interpretation emphasized the magnetically gapped surface and associated its
response with the continuum parity anomaly. The lattice picture developed
here suggests a different assignment. A fully gapped two-dimensional lattice
band cannot by itself carry a standalone half-integer dc Hall conductance.
Instead, the essential low-energy degree of freedom is the opposite,
gapless surface Dirac cone. Its Fermi surface lies in a locally
parity-symmetric regime, while the magnetically gapped surface together with
the three-dimensional bulk provides the parity-breaking high-energy
completion required for an effectively single Dirac fermion
\cite{ZouPRB2022,ZouPRB2023,ShenPRB2026}.

\begin{figure}[p]
  \centering
  \includegraphics[angle=-90,width=0.98\linewidth]{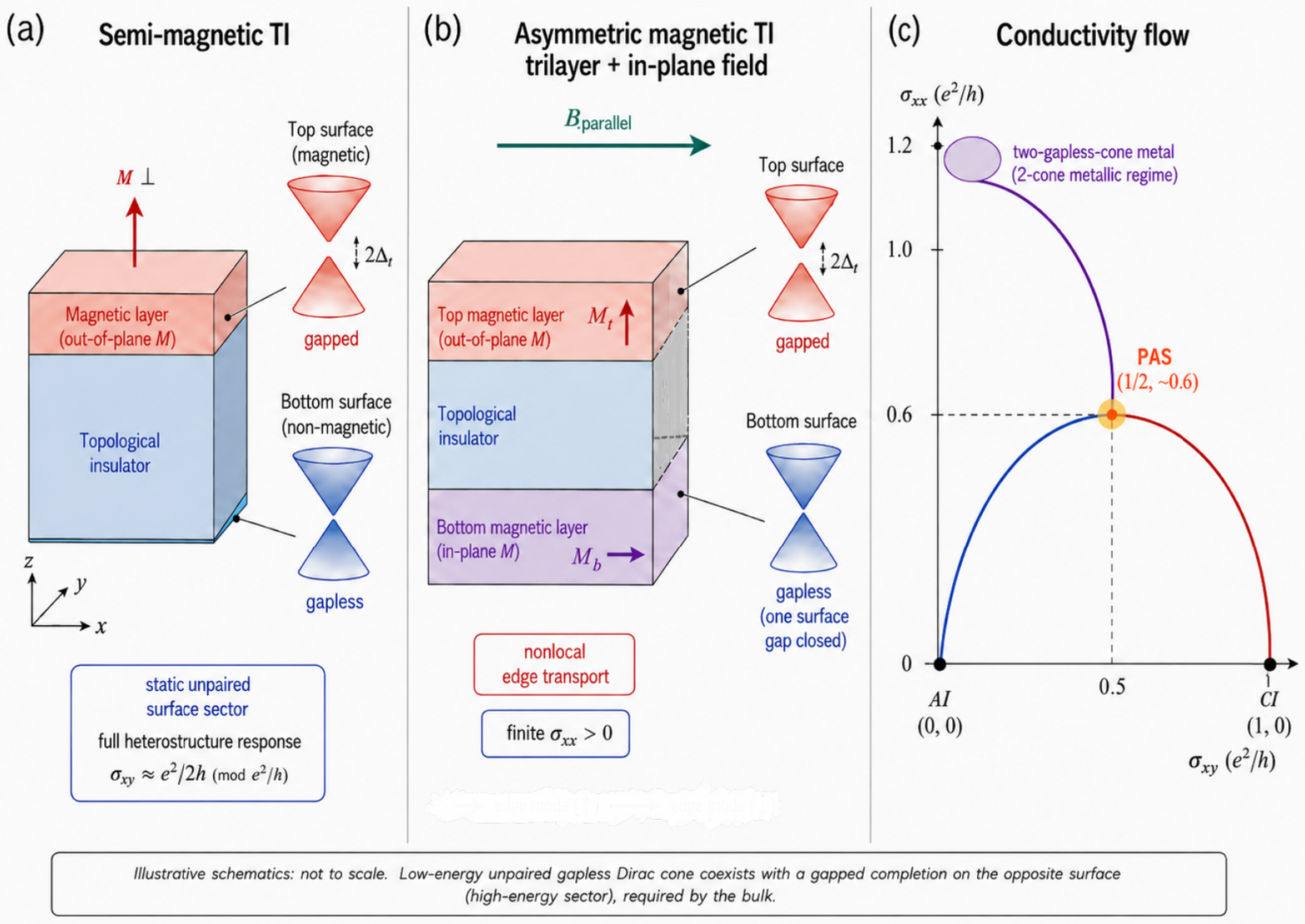}
  \caption{
  From a static unpaired surface Dirac cone to a field-tuned
  parity-anomalous semimetal.
  (a) In the semi-magnetic topological insulator of Mogi \emph{et al.},
  one surface is magnetically gapped while the opposite surface remains
  gapless. In the present interpretation, the half response belongs to the
  complete heterostructure containing the gapless low-energy cone and its
  high-energy completion.
  (b) In an asymmetric magnetic topological-insulator sandwich, an in-plane
  field can close one surface Dirac gap while the opposite surface remains
  gapped, thereby producing an effectively unpaired massless Dirac cone.
  (c) Schematic conductivity flow: trajectories originating from Chern- or
  axion-insulator states approach the PAS near
  $(1/2,\sim0.6)e^2/h$. When the second surface gap also closes, the flow
  continues toward a two-gapless-cone metallic regime near
  $(0,\sim1.2)e^2/h$. The curves are illustrative rather than reproductions
  of experimental data.
  }
  \label{fig:experiment-story}
\end{figure}

This interpretation also changes the meaning of the accompanying
longitudinal transport. A nonzero longitudinal conductivity is not merely an
imperfection superposed on an otherwise ideal half-quantized Hall insulator.
It is a natural consequence of the gapless Dirac sector. The experimentally
measured Hall response belongs to the complete heterostructure, not to a
separately addressable ``half Hall layer.'' The decomposition into a gapless
low-energy sector and a gapped high-energy completion is therefore a
theoretical interpretation of the full band structure and its symmetry,
rather than a layer-resolved measurement. The semi-magnetic geometry is
illustrated schematically in Fig.~\ref{fig:experiment-story}(a).

A more revealing realization is obtained when the single gapless cone is
created continuously. In asymmetric magnetic topological-insulator
sandwiches, the two magnetic layers can have substantially different
anisotropy fields. An in-plane magnetic field can therefore rotate one
surface magnetization into the plane and close its exchange gap while the
opposite surface remains magnetically gapped. With increasing field, the
system passes through an intermediate regime containing one gapless and one
gapped surface Dirac cone before the second surface gap eventually closes.
This provides a direct way to create, and then remove, an effectively
unpaired massless Dirac fermion. The Chern-insulator and axion-insulator starting states themselves are standard magnetic-topological-insulator phases \cite{Yu2010,Chang2013,Xiao2018}.

Recent field-tuned experiments make this evolution visible in transport
\cite{YangAdvMat2026,WangPRL2026,ZhuoPRL2026}. Starting from either a
Chern-insulator or an axion-insulator configuration, the conductivity flow
approaches an intermediate regime near
$(\sigma_{xy},\sigma_{xx})\simeq(1/2,0.6)e^2/h$ when the first surface gap
closes \cite{WangPRL2026}. When the second surface subsequently becomes
gapless, the flow continues toward a two-cone metallic regime with vanishing
Hall response and a longitudinal conductivity of order $1.2e^2/h$. The
important observation is therefore not the numerical value
$\sigma_{xy}=e^2/2h$ by itself, but the existence of a distinct metallic
intermediate state associated with only one gapless surface cone. 
This two-stage evolution sharply distinguishes the parity-anomalous
semimetal from an ordinary integer-quantum-Hall transition. In the latter,
$\sigma_{xy}=e^2/2h$ marks an unstable critical point separating integer
Hall plateaus. Here the half-Hall response occurs together with a finite
longitudinal conductivity and is approached from different insulating
initial states as a stable intermediate regime. 

The magnetic field provides an additional symmetry control. An in-plane
field can close the exchange gap without directly generating the mass term
that destroys the low-energy parity symmetry, whereas a sufficiently strong
perpendicular component restores a surface mass and drives the system toward
Chern- or axion-insulator fixed points. This behavior is consistent with the
central criterion of the present Perspective: the half response survives
while the Fermi surface remains within the locally parity-symmetric Dirac
sector and disappears when that sector is driven outside the symmetry-
preserving window. The corresponding surface configurations and
conductivity flow are summarized in Fig.~\ref{fig:experiment-story}(b,c).

This distinction is important. A transport experiment probes the entire
device, including the gapless surface and side-surface states. It does not
separately measure a half-quantized Hall conductance localized on the
remaining massive surface. Indeed, numerical calculations associated with
the experiment find that the boundary current is constructed from metallic
modes of massless Dirac electrons. The observation of boundary-sensitive
chiral transport is therefore consistent with the parity-anomalous-semimetal
picture, but it does not by itself establish a standalone $C=1/2$ insulating
state associated with a massive Dirac surface.

Disorder determines how metallic this state is without changing its
topological origin. A recent transport theory of unpaired Dirac fermions,
including self-consistent disorder broadening, vertex corrections, and
quantum interference, finds a nearly semi-elliptic relation between
$\sigma_{xy}$ and $\sigma_{xx}$ \cite{FuScaling2026}. The PAS appears as
the fixed point with half-quantized Hall conductivity and a finite minimum
longitudinal conductivity $\sigma^*$ whose value depends on disorder.
Topology and scattering therefore play different roles: the former fixes the
half-Hall background, whereas the latter controls the longitudinal
conductivity of the gapless Dirac metal. The experimentally observed value
near $0.6e^2/h$ is consequently not a second topological quantum but a
characteristic conductivity of the realized metallic state.

Viewed in this way, the experiments are more than successive realizations
of a half-quantized Hall plateau. They expose the structure of a single Dirac
fermion on a lattice. The semi-magnetic heterostructure realizes the unpaired
surface sector statically; the in-plane-field experiments create and remove
the same sector dynamically. In both cases the anomalous response emerges
from the coexistence of a finite low-energy region with restored parity
symmetry and a high-energy parity-breaking completion. The experiments
therefore provide a direct materials setting in which the ultraviolet
regularization of the parity anomaly becomes part of the observable band
structure rather than an abstract prescription.

\section{A unified lattice perspective}

The chiral and parity anomalies look different when presented 
through their standard field-theory calculations. 
The chiral anomaly is formulated as nonconservation of an axial current, 
whereas the parity anomaly is formulated through a parity-odd effective action or half-quantized Hall coefficient. 
On a lattice, however, their microscopic architectures are strikingly similar.

\begin{center}
\begin{tabular}{p{0.24\linewidth}p{0.32\linewidth}p{0.32\linewidth}}
\toprule
 & \textbf{Chiral anomaly} & \textbf{Parity anomaly} \\
\midrule
Dimension & $1+1$ and $3+1$ & $2+1$ \\
Low-energy object & massless Dirac fermion / LLL channel & single massless Dirac cone \\
Local symmetry & chiral symmetry at the Fermi surface & mirror (parity) symmetry at the Fermi surface \\
High-energy role & chiral-symmetry-breaking completion & parity-breaking completion \\
Observable response & spectral flow, axial pumping, $\mathbf E\cdot\mathbf B$ response & half Hall response with finite $\sigma_{xx}$ \\
Material realization & Chern-insulator boundary; Weyl/Dirac systems & parity-anomalous semimetal on TI surfaces \\
\bottomrule
\end{tabular}
\end{center}

Three concepts should remain distinct. \emph{Explicit symmetry breaking} is a term in the Hamiltonian and by itself produces no anomaly. \emph{Topology} gives the global consistency conditions for the complete bands and determines, for example, integer Hall conductance in a gapped two-dimensional lattice insulator. The \emph{anomaly} is the infrared--ultraviolet mismatch that appears when a low-energy single-fermion theory has a symmetry that cannot be maintained in the same form by its ultraviolet completion.

\begin{figure}[p]
  \centering
  \includegraphics[angle=-90, width=0.94\linewidth]{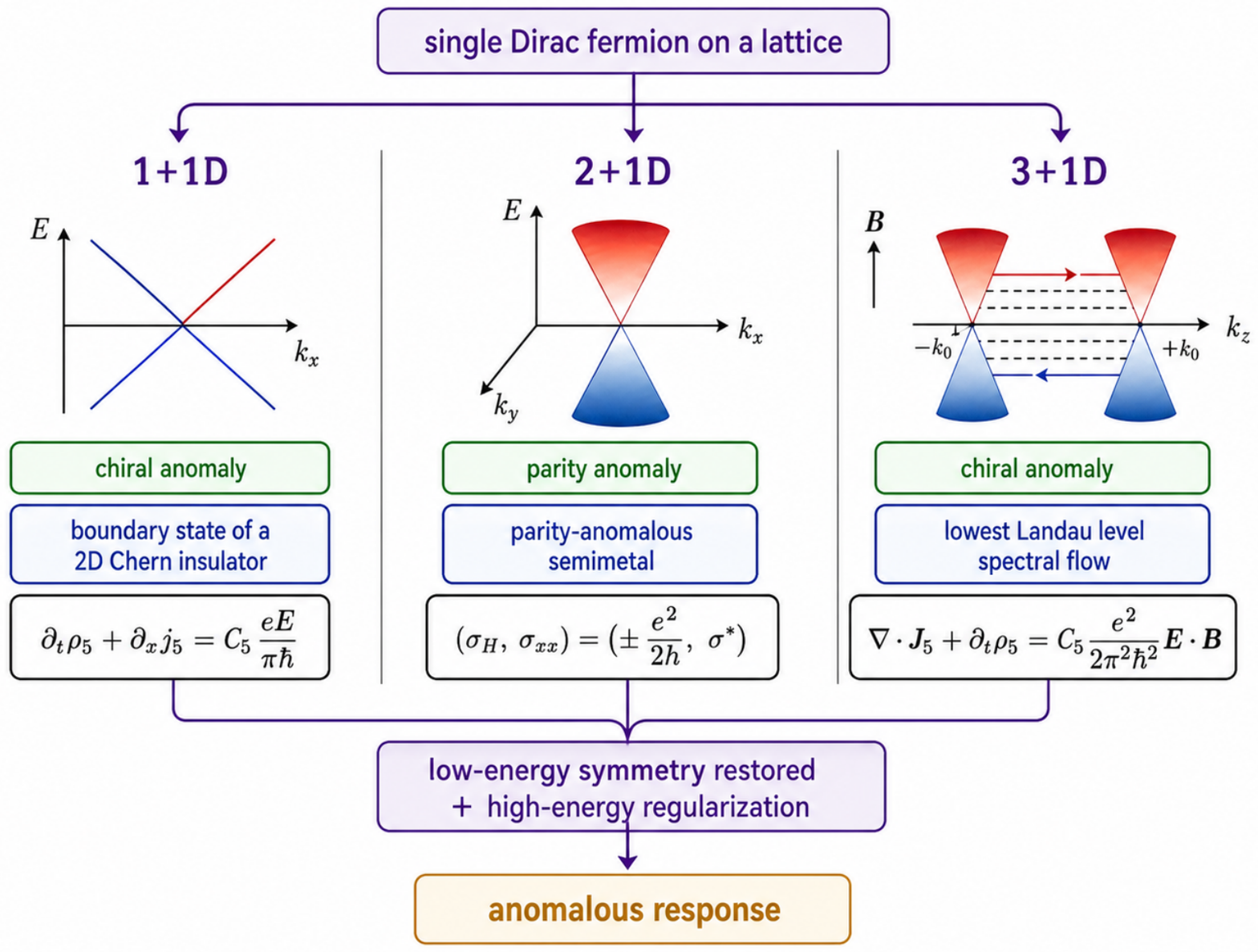}
  \caption{Unified lattice view. In $1+1$ and $3+1$ dimensions the low-energy symmetry is chiral and the anomaly appears through spectral flow; in $2+1$ dimensions the low-energy symmetry is parity and the anomalous state is a half-Hall semimetal. In each case the full response requires both the locally symmetric Dirac sector and its high-energy lattice completion.}
  \label{fig:unified-map}
\end{figure}

This separation resolves several apparent paradoxes. In the chiral problem the high-energy states violate chiral symmetry, yet the anomaly coefficient is quantized when the Fermi surface remains locally chiral. In the parity problem a mass term breaks mirror symmetry and therefore allows Hall response, yet a massive lattice insulator is integer quantized. The half Hall response appears when the low-energy cone is massless and parity symmetric while the necessary symmetry breaking is displaced to high energy. The common architecture of the two anomalies is summarized in Fig.~\ref{fig:unified-map}.

This is also where the comparison with lattice gauge theory is most useful. Wilson, domain-wall, and related lattice fermions already teach us that anomaly, doubling, and ultraviolet regularization are inseparable. The condensed-matter contribution is not a new solution of the fermion-doubling problem. It is to turn that regulator structure into band physics: the Brillouin zone is physical, the symmetry-restoring region can occupy a finite window around the Fermi surface, and the ultraviolet completion can be interrogated through spectroscopy and transport rather than removed by a continuum limit.

\begin{tcolorbox}
  [colback=boxblue,colframe=blue!45!black,title={Models for single Dirac fermions},
  fonttitle=\bfseries]
  For the continuous case, 
  \begin{equation}
  \mathcal{H} = vk_x \sigma_x + vk_y \sigma_y +\chi \Theta[-M(\mathbf{k})] M(\mathbf{k}) \sigma_z
  \end{equation}
  where $v$ is the effective velocity,$\sigma_{x,y,z}$ are the Pauli matrices, 
  $M(\mathbf{k})=m_0-b(k_x^2+k_y^2)$ ($m_0\geq 0,b>0$) is a function that vanishes at $k_c=\sqrt{b/m_0}$, 
  $\chi=\pm 1$ represents the chirality, and $\Theta$ is the Heaviside step function. 
  For the lattice case,
  \begin{equation}  
  \mathcal{H} =  t\sin(k_x a)\sigma_x + t\sin(k_y a)\sigma_y 
  +\chi \Theta[-M(\mathbf{k})]M(\mathbf{k})\sigma_z
\end{equation}
  where $M(\mathbf{k})=m_0-b(2-\cos(k_x a)-\cos(k_y a))$ is the lattice version of the mass term, 
  and $a$ is the lattice constant. The Heaviside function $\Theta[-M(\mathbf{k})]$ ensures 
  that the mass term is only active outside the low-energy window, 
  allowing for a single massless Dirac cone at low energies 
  while providing a high-energy completion that breaks the symmetry. The energy dispersions are continuous and differentiable.
  For one-dimensional case, the model is obtained by taking $k_y=0$, while 
  for three-dimensional case, the model can be constructed by replacing 
  the Pauli matrices with the appropriate $4\times 4$ Dirac matrices 
  and including the $k_z$ term. Generally speaking, the $d$-dimensional model can be constructed similarly 
  by solving $d+1$-dimensional Dirac systems with finite thickness 
  \cite{ZouPRB2023,WangFuShen2025,ShenSpringer2012}.
\end{tcolorbox}

\section{Outlook and conclusion}

The lattice viewpoint suggests several experimental tests that go beyond observing a familiar coefficient. The first is to map the boundary of the local-symmetry window. In the chiral case, gating a single-cone system until the Fermi surface leaves the chiral-symmetric region should drive the coefficient $C_5$ away from its quantized value. In the parity case, the half Hall response should persist only while the Fermi surface remains inside the mirror-symmetric gapless sector. Combining transport with ARPES, quantum oscillations, or layer-sensitive spectroscopy would test this connection directly.

A second direction is to tune the ultraviolet completion while leaving the low-energy cone nearly unchanged. Magnetic anisotropy, heterostructure thickness, surface hybridization, strain, and interface exchange all change how a Dirac surface state merges into high-energy bands. If the present picture is correct, such modifications can alter the crossover away from the anomalous regime without changing the protected response as long as the Fermi surface remains inside the local-symmetry window. This provides a falsifiable distinction between a genuinely local-symmetry-protected anomaly and an accidental half value produced by a particular band parameter.

A third lesson is methodological. Negative longitudinal magnetoresistance is not unique evidence for the chiral anomaly, and a half Hall coefficient by itself is not unique evidence for the parity anomaly. The anomaly should be identified by a correlated set of observations: the low-energy symmetry, the single-cone band structure, the high-energy completion, and the expected longitudinal as well as transverse response.

The main conclusion is therefore not that condensed matter simply reproduces a field-theory anomaly. It makes the regularization visible. A single Dirac fermion on a lattice is possible because the low-energy sector can recover a symmetry that the full high-energy band structure does not preserve. In $1+1$ and $3+1$ dimensions this structure appears as chiral spectral flow; in $2+1$ dimensions it produces the parity-anomalous semimetal. The finite Brillouin zone is not merely a cutoff on the continuum theory. It is the physical stage on which the infrared symmetry and ultraviolet completion meet.

\printbibliography[title={References}]

\end{document}